\documentclass[conference]{IEEEtran}
\IEEEoverridecommandlockouts
\usepackage{cite}
\usepackage{amsmath,amssymb,amsfonts}
\usepackage{algorithmic}
\usepackage{graphicx}
\usepackage{textcomp}
\usepackage{xcolor}
\usepackage{booktabs}
\usepackage{pifont}
\usepackage{multirow}
\usepackage{adjustbox}
\usepackage{hyperref}
\def\BibTeX{{\rm B\kern-.05em{\sc i\kern-.025em b}\kern-.08em
    T\kern-.1667em\lower.7ex\hbox{E}\kern-.125emX}}
\begin{document}

\title{StereoSync: Spatially-Aware Stereo \\Audio Generation from Video\\
\thanks{* Equal contribution.\\Corresponding author's email: \text{christian.marinoni@uniroma1.it}}
}
\author{\IEEEauthorblockN{Christian~Marinoni$^{\flat*}$, Riccardo~F.~Gramaccioni$^{\flat*}$, Kazuki~Shimada$^{\sharp}$,\\Takashi~Shibuya$^{\sharp}$, Yuki~Mitsufuji$^{\sharp \natural}$, and~Danilo~Comminiello$^{\flat}$}\\
    \IEEEauthorblockN{\textit{$^{\flat}$Sapienza University of Rome, Italy}\\\textit{$^{\sharp}$Sony AI, Tokyo, Japan}\\\textit{$^{\natural}$Sony Group Corporation, Tokyo, Japan}
}}

\maketitle

\begin{abstract}

Although audio generation has been widely studied over recent years, video-aligned audio generation still remains a relatively unexplored frontier. 
To address this gap, we introduce StereoSync, a novel and efficient model designed to generate audio that is both temporally synchronized with a reference video and spatially aligned with its visual context. Moreover, StereoSync also achieves efficiency by leveraging pretrained foundation models, reducing the need for extensive training while maintaining high-quality synthesis. 
%
Unlike existing methods that primarily focus on temporal synchronization, StereoSync introduces a significant advancement by incorporating spatial awareness into video-aligned audio generation. Indeed, given an input video, our approach extracts spatial cues from depth maps and bounding boxes, using them as cross-attention conditioning in a diffusion-based audio generation model. Such an approach allows StereoSync to go beyond simple synchronization, producing stereo audio that dynamically adapts to the spatial structure and movement of a video scene. 
We evaluate StereoSync on \textit{Walking The Maps}, a curated dataset comprising videos from video games that feature animated characters walking through diverse environments. Experimental results demonstrate the ability of StereoSync to achieve both temporal and spatial alignment, advancing the state of the art in video-to-audio generation and resulting in a significantly more immersive and realistic audio experience.
\end{abstract}


\section{Introduction}
Video-to-audio (V2A) generation, the task of synthesizing realistic sound from visual content, has emerged as a crucial technology with applications spanning cinema post-production, video game sound design, and immersive multimedia experiences. While recent advances in foundation models have drastically improved the quality of generated audio \cite{Liu2023AudioLDM2L, Huang2023MakeAnAudioTG}, creating spatially-aware sounds that match the visual environment remains a significant challenge.

Recent research in V2A has primarily focused on semantic alignment, ensuring that generated audio matches the content and context of the visual scene \cite{Sheffer2022IHY, chen2024images, Gramaccioni2024L3DAS23L3, Gan2020FoleyML}. More recent approaches have begun to address temporal synchronization, generating audio that follows the precise timing of visual events \cite{chen2020GeneratingVA, 9782577, 9126216}. However, a crucial third dimension remains largely unexplored: spatial awareness. Current 
reliance of systems on monophonic output and lack of spatial understanding fundamentally limits the immersive potential of the generated audio, thus resulting in sounds that feel artificially disconnected from the spatial layout of the visual scene.

\begin{figure}[t!]
    \centering
    \includegraphics[width=0.85\linewidth]{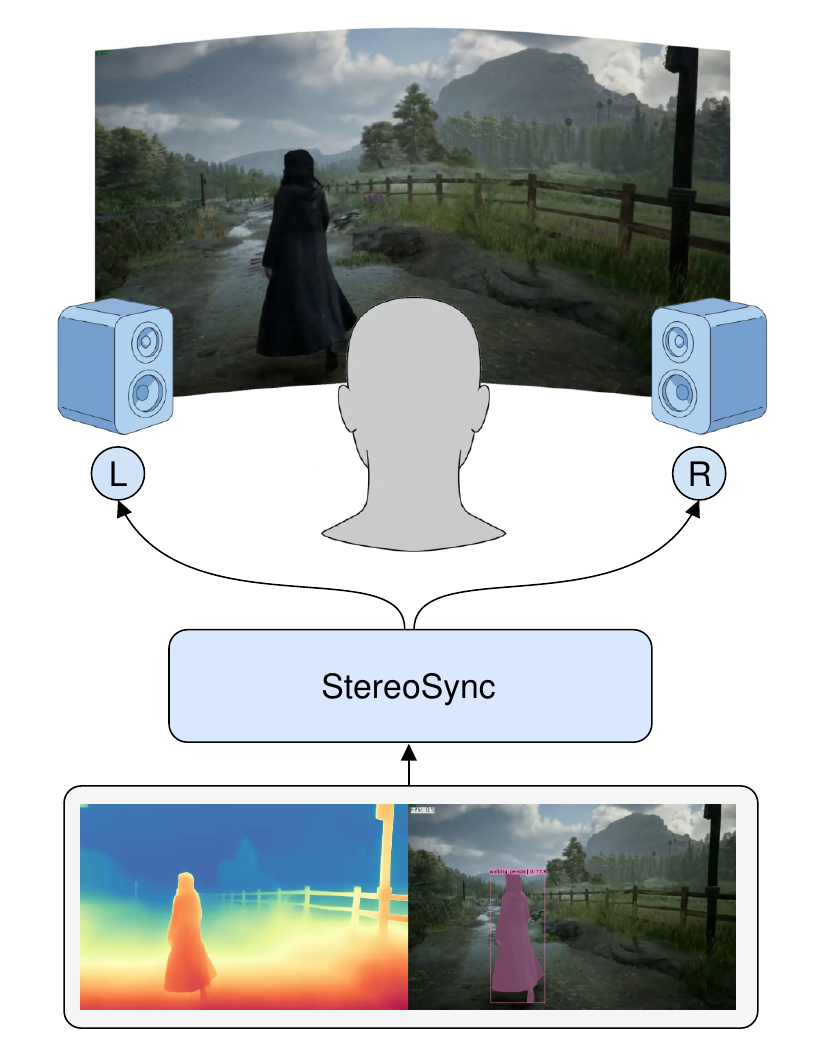}
    \caption{StereoSync generates a stereo audio that resembles the spatial context of an input video.}
    \label{fig:waveform}
\end{figure}

Stereo audio generation offers a path toward more immersive experiences by incorporating spatial cues that mirror the physical arrangement of sound sources. However, existing stereo generation methods typically rely on extensive training with specialized datasets and complex architectures dedicated to spatial learning \cite{zhou2020sep, Kleijn2023MultiChannelAS, Park2024ImagetoStereoAG}. This high computational cost has limited broader adoption and hindered further investigation on spatially-aware audio generation. To overcome this limitation, in this paper, we leverage and efficiently adapt foundation models to tackle the challenge of spatially-aware audio generation, thus reducing the need for extensive training while maintaining high-quality results.

In particular, we present StereoSync, a novel approach that relies on pretrained foundation models and minimal fine-tuning to generate spatially-enriched stereo audio. Our framework processes input videos through multiple specialized pathways: extracting depth maps to capture scene geometry, computing bounding boxes to track moving subjects (in our case, people, though the approach generalizes to other objects). These features are processed through dedicated encoders
, and then utilized as cross-attention conditioning signals in a diffusion-based generation framework. By combining these spatial cues with semantic information derived from audio samples and temporal alignment from waveform analysis, StereoSync creates audio that maintains coherent spatial relationships while ensuring semantic fidelity and temporal synchronization with the visual content. This efficient adaptation of foundation models not only reduces computational requirements but also achieves state-of-the-art performance in generating audio that is simultaneously semantically meaningful, temporally precise, and spatially aware. The task handled by the proposed model is outlined in Fig.~\ref{fig:waveform}.

The key contributions of our work are:
\begin{itemize}
\item \textbf{Advancing spatial awareness in V2A generation}. We tackle the underexplored dimension of spatial awareness in video-to-audio generation by incorporating stereo audio that represents the physical layout of sound sources in the visual scene, enhancing the perceived immersive experience.
\item \textbf{Multimodal feature extraction for audio generation}. Our framework processes diverse visual and temporal features to guide audio synthesis, including depth maps for scene geometry, bounding boxes for object tracking, and waveform envelopes for temporal dynamics, to inform the audio generation process. These features are integrated into a diffusion-based generation model through cross-attention conditioning, ensuring a coherent alignment between audio and video content.
\item \textbf{Efficient adaptation of foundation models}. By adapting pretrained foundation models with minimal fine-tuning, StereoSync achieves improved spatially-aware audio synthesis while significantly reducing computational requirements.
\end{itemize}

The rest of the paper is organized as follows. Section~\ref{sec:works} presents the related works, Section~\ref{sec:method} the proposed method, while in Section~\ref{sec:exp} we discuss the experimental results and in Section~\ref{sec:res} we validate the obtained results. Finally, conclusions are drawn in Section~\ref{sec:con}.

\section{Related Works}
\label{sec:works}

\textbf{Video-to-Audio Generation.} Video-to-audio synthesis has emerged as a critical research area in multimedia production, driven by advancements in deep learning. Early efforts focused on aligning visual and audio modalities. Pioneering work like Im2Wav \cite{Sheffer2022IHY} leveraged CLIP-based visual features \cite{Radford2021LearningTV} to guide audio generation, while RegNet \cite{chen2020GeneratingVA} explored the use of GANs in conjunction with video encoders to improve temporal correspondence between video and audio. However, these early methods were limited to generating mono-audio, restricting their ability to capture the spatial dynamics inherent in visual scenes.

The field subsequently evolved to address increasingly sophisticated aspects of audio-visual alignment. SpecVQGAN \cite{SpecVQGAN_Iashin_2021} marked a significant advance by introducing real-time audio generation capabilities on consumer hardware, leveraging a VQGAN-based codebook architecture and an innovative window-based GAN for efficient spectrogram-to-waveform conversion. Building on this foundation, Diff-Foley \cite{NEURIPS2023_98c50f47} introduced a novel approach combining latent diffusion models with contrastive audio-visual pretraining (CAVP), demonstrating superior temporal and semantic alignment while maintaining strong generalization capabilities through downstream fine-tuning.
The conditional generation of Foley sounds saw further advancement with CondFoleyGen \cite{du2023conditional}, which tackled the challenge of generating soundtracks for silent videos while matching user-provided audio examples. Their two-stage approach, combining video-based sound prediction with style-aware generation, proved effective through human evaluation and objective metrics. 

However, these methods lack human-intelligible controls, limiting their utility in practical sound design applications \cite{gramaccioni2024folai}.
SyncFusion \cite{Comunit2023SyncfusionMO} introduced a practical framework for sound designers, automatically extracting action onsets from videos to guide synchronized sound effect generation through diffusion models, while preserving creative control through onset tracks and conditioning embeddings. 
More recently, T-Foley \cite{Chung2024TFoleyAC} achieved precise temporal alignment through a novel temporal event feature and Block-FiLM conditioning mechanism, demonstrating particular success in applications like vocal mimicry. 

Despite the progress made, all these approaches share a common limitation: they generate monophonic outputs that fail to represent the spatial distribution and dynamics of the actions depicted in the input video, resulting in a lack of immersiveness and accuracy in the modelling of sound directionality.

\textbf{Stereo audio generation.} While video-to-audio generation has primarily focused on monophonic output, multiple recent works have explored the challenge of stereo audio synthesis, even though primarily in the context of mono-to-stereo conversion. Parametric stereo generation \cite{DBLP:conf/ismir/SerraSPAPBC23} pioneered a deep learning approach to this task, introducing generative models that predict parametric stereo parameters through autoregressive and masked token modeling. This work demonstrated that generative approaches could outperform traditional decorrelation methods, though its scope was limited to converting existing mono signals.

MusicHiFi \cite{Zhu2024MusicHiFiFH} advanced the field by introducing a cascade of GANs designed to generate high-fidelity stereophonic audio from mel-spectrograms. Their unified architecture, incorporating both bandwidth extension and mono-to-stereo conversion modules, achieved significant improvements in audio quality and spatialization control. The BEWO \cite{sun2024both} framework further expanded the possibilities of spatial audio generation by introducing SpatialSonic, the first model capable of generating controllable spatial audio through spatial-aware encoders and azimuth state matrices, supported by their novel BEWO-1M dataset.
On the other hand, MAFNet \cite{zang2021multiatt} made strides in incorporating visual information into stereo generation through a multi-attention fusion network, achieving competitive results on video datasets like FAIR-Play \cite{gao2019visualsound} and YT-MUSIC \cite{morgado2018self}. However, their approach still relies on converting an existing mono signal to stereo, using video frames primarily as a reference rather than as the primary source for audio generation.

Therefore, a fundamental limitation of existing approaches is their treatment of stereo generation as a post-processing step, focusing on converting mono signals to stereo rather than generating spatially-aware stereo audio directly from visual content. Our work addresses this limitation by approaching stereo generation as an integral part of the video-to-audio pipeline, leveraging foundation models to generate spatially-aware stereo audio directly from visual content.



\section{StereoSync}
\label{sec:method}

\subsection{Step 1: extracting relevant features from videos}

\begin{figure*}
    \centering
    \includegraphics[width=\linewidth]{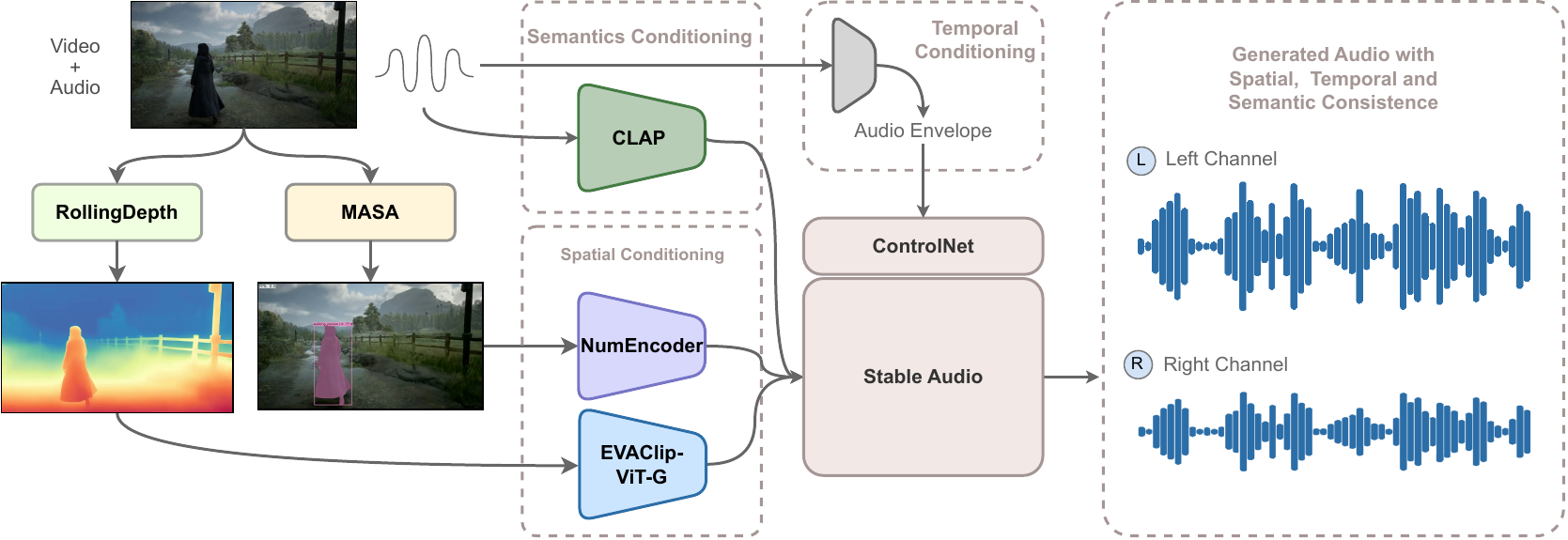}
    \caption{\textbf{StereoSync architecture}: the proposed method extracts depth maps and bounding boxes from input videos through video foundation models RollingDepth and MASA. 
    Relevant features are then extracted to represent the spatiality of the scene. These features, along with a CLAP embedding of the audio sample that is expected to characterize the semantics of the final audio, are used to condition the audio synthesis model. 
    Temporal control is provided through an envelope signal which is used as ControlNet input. 
    The only weights that are trained in the architecture are the ControlNet weights and projection layers weights used to map the conditioning embeddings to the correct shape required by Stable Audio, making our model lightweight and efficient.}
    \label{fig:stereosync-architecture}
\end{figure*}

The first step in our pipeline involves extracting rich spatial information from the input video frames. We leverage two state-of-the-art foundation models to obtain complementary spatial features: depth maps that capture the overall scene geometry and precise bounding boxes that track moving subjects.
To extract accurate and temporally consistent depth information, we employ RollingDepth \cite{ke2024video}, a recent advancement in video depth estimation that builds upon latent diffusion models. Unlike traditional approaches that process frames independently, RollingDepth analyzes frame triplets and uses an optimization-based registration algorithm to ensure temporal consistency across the entire video sequence. This approach is particularly valuable for our task as it provides stable depth estimates even under camera motion and varying depth ranges, enabling more reliable spatial audio generation.

For tracking moving subjects within the scene, we utilize MASA \cite{masa}, a robust instance association framework that exploits the Segment Anything Model (SAM) \cite{kirillov2023segment} for zero-shot object tracking. The ability of MASA to generalize across diverse domains without requiring domain-specific training makes it particularly suitable for our application. The framework provides precise bounding box coordinates for tracked subjects across frames, maintaining consistent object identity even in complex scenes. While our current implementation focuses on tracking people, MASA's universal adapter architecture allows for tracking any class of objects detected by foundation models.

The combination of these two feature extraction methods provides our model with:
(1) Global spatial context through dense depth maps; (2) Local spatial information through precise object tracking.

\subsection{Audio Synthesis Model}

Our audio synthesis model utilizes Stable Audio Open \cite{Evans2024StableAO}, a state-of-the-art latent diffusion model (LDM) for generating high-quality, stereo audio at 44.1 kHz. While Stable Audio excels at generating semantically rich audio from text prompts, it lacks explicit mechanisms for temporal and multimodal conditioning, making it unsuitable for video-to-audio (V2A) tasks. To address this limitation, we introduce novel conditioning strategies leveraging multimodal embeddings and temporal guidance derived directly from the input video and audio.

\subsubsection{Spatial Control} Visual embeddings are derived from the input video to incorporate spatial context. More formally, a video sample has size $\mathbf{v} \in \mathbb{R}^{T \times C \times H \times W}$, where $T=60$ is the number of frames for 2 seconds long video chunks at 30 \textit{fps}, $C=3$ are the RGB channels, $W=1280$ and $H=720$ are the width and height of the frames. From each video we extract bounding boxes with MASA and depth maps with RollingDepth. Bounding boxes are computed at the same frame rate as the video (i.e. 30 $fps$) to ensure tracking even for fast actions, while depth maps, which provide a global spatial context, are calculated over fewer frames. Indeed, we extract depth maps at 4 \textit{fps}, producing 8 depth frames for each 2 second video chunk processed. 

We then retrieve relevant features from both bounding boxes and depth maps. Specifically:
\begin{itemize}
\item Depth Maps are processed frame-by-frame using the EVAClip-ViT-G \cite{Sun2023EVACLIPIT} encoder; these maps capture the scene geometry and provide spatial cues for the soundscape. To match the size required by  EVAClip-ViT-G, we resize each frame of the depth maps to the size $344 \times 256$. For each of the 8 depth frames, the encoder outputs an embedding of size 768. Consequently, the embedding produced for each depth map has dimension [8, 768], which is then projected to dimension [8, 512] by a linear projector to match the dimensions required by Stable Audio.
\item Bounding Boxes are derived using MASA and encoded with the NumberConditioner of Stable Audio Open, which is designed for encoding lists of floats. Each bounding box for a video frame is represented as a quadruplet of floats $[x_i, y_i, x_f, y_f]$, where $i$ denotes the initial pixel and $f$ the final pixel in both the $x$ and $y$ directions. From these quadruplets, we extract four separate sequences, each corresponding to one of the components (e.g., a list of all $x_i$ values across frames). Each of these sequences is then individually encoded using the NumberConditioner.

\end{itemize}

\subsubsection{Semantic Control} Our approach uses audio embeddings obtained via the CLAP \cite{laionclap2023} encoder to define the semantic features that the generated audio must possess.

Both semantic and spatial information embeddings are integrated into the diffusion process as cross-attention conditioning signals, enabling the generation of spatially and semantically aligned stereo audio.

\subsubsection{Temporal Control} Temporal alignment is guided by envelopes extracted directly from the ground truth audio using the Librosa library \footnote{\url{https://librosa.org/doc/main/generated/librosa.feature.rms.html}}. The $i$-th sample of the temporal sequence representing the envelope is then calculated on a window of the audio signal $\mathbf{y}$ as follows:
\begin{equation}
    \mathbf{r}_i = \mathbf{RMS}_i(\mathbf{y)} = \sqrt{\frac{1}{W} \sum_{t=ih}^{ih+W}\mathbf{y}^{2}(t)},
\label{eqn:RMS}
\end{equation}
where  $W=512$ is the window size and $h=128$ is the hop size.

The envelope signal is processed using a ControlNet-inspired architecture \cite{chen2024pixartalpha}, which allows fine-grained temporal adjustments during audio generation.

Since the input of the ControlNet must be the same size as the input to the diffusion process, we interpolate it to match the original audio size, thus obtaining $\mathbf{r} \in \mathbb{R}^{Ch \times L}$, where $Ch$ represents the number of audio channels, that is 2 for stereo signals, and $L$ is the time duration of the audio expressed in samples, i.e. 88200 for 2 second audio at 44.1 kHz.

\subsubsection{Diffusion Process} our audio model is based on Stable Audio and follows the standard latent denoising diffusion formulation. Given a noisy latent representation $\mathbf{z} = \mathcal{E}(\mathbf{y})$, obtained from $\mathbf{y} \in \mathbb{R}^{Ch \times L}$ (where $Ch$ and $L$ are the same as defined before), at time step $t$, the model learns to estimate the noise $\epsilon_{\theta}(\mathbf{z}_t, t, \mathbf{F}, \mathbf{r})$ conditioned on a set of spatial and semantic embeddings $\mathbf{F}={\mathbf{f}_1, \mathbf{f}_2,..., \mathbf{f}_n}$ and the temporal control signal $\mathbf{r}$. 
In the forward process, Gaussian noise is slowly added to the original data distribution with a fixed schedule $\alpha_1, \ldots, \alpha_T$, where $T$ is the total timesteps, and $\bar{\alpha}_t = \prod_{i=1}^{t} \alpha_i$:

\begin{equation}
q(\mathbf{z}_t | \mathbf{z}_{t-1}) = \mathcal{N}(\mathbf{z}_t; \sqrt{\alpha_t} \mathbf{z}_{t-1}, (1 - \alpha_t) \mathbf{I})
\end{equation}
\begin{equation}
q(\mathbf{z}_t | \mathbf{z}_0) = \mathcal{N}(\mathbf{z}_t; \sqrt{\bar{\alpha}_t} \mathbf{z}_0, (1 - \bar{\alpha}_t) \mathbf{I}).
\end{equation}



The audio diffusion model is trained on the same L2 loss on which Stable Audio models are trained \cite{Evans2024LongformMG}.

After training, LDMs generate latents by sampling through the reverse process with $\mathbf{z}_T \sim \mathcal{N}(0, \mathbf{I})$ formulated as:

\begin{equation}
p_\theta(\mathbf{z}_{t-1} | \mathbf{z}_t) = \mathcal{N}(\mathbf{z}_{t-1}; \mu_\theta(\mathbf{z}_t, t, \mathbf{F}, \mathbf{r}), \sigma_t^2 \mathbf{I})
\end{equation}

\begin{equation}
\mu_\theta(\mathbf{z}_t, t, \mathbf{F}, \mathbf{r}) = \frac{1}{\sqrt{\alpha_t}} \left( \mathbf{z}_t - \frac{1 - \alpha_t}{\sqrt{1 - \bar{\alpha}_t}} \epsilon_\theta(\mathbf{z}_t, t, \mathbf{F}, \mathbf{r}) \right)
\end{equation}

\begin{equation}
\sigma_t^2 = \frac{1 - \bar{\alpha}_{t-1}}{1 - \bar{\alpha}_t} (1 - \alpha_t).
\end{equation}

Finally, the desired output  $\hat{\mathbf{y}}$ is obtained by decoding the generated latent $\mathbf{z}_0$ with a decoder $\mathcal{D}$.

We freeze the pre-trained weights of the diffusion model and only train the ControlNet layers, which process the envelope, and the linear projections that align CLAP, bounding boxes and depth maps embeddings to the conditioning dimensions of Stable Audio. 
The ControlNet block is trained with the v-prediction MSE loss $\mathcal{L} = \mathbb{E}[v_\theta(\mathbf{z_t}, t, \mathbf{r_c}) - v]$, where $v = \sqrt{\bar{\alpha_t}}\epsilon - \sqrt{1-\bar{\alpha_t}} x_0$.
By jointly leveraging bounding boxes and depth maps embeddings for spatial control, CLAP audio embedding for semantic control and the ControlNet mechanism for temporal alignment, our model generates waveforms that are spatially, semantically and temporally aligned with the reference input video. A block diagram of the proposed architecture is shown in Fig.~\ref{fig:stereosync-architecture}.

\section{Experiments}
\label{sec:exp}
\subsection{Dataset}

The \textit{Walking The Maps} introduced in \cite{gramaccioni2024folai} dataset is specifically designed for evaluating video-to-audio (V2A) models in the context of generating footsteps sounds, a common and challenging task in sound design for audiovisual works. The dataset addresses the lack of publicly available resources suitable for V2A tasks, which require high-resolution video paired with sound design-quality audio.

\begin{figure}[t!]
    \centering
    \includegraphics[width=\linewidth]{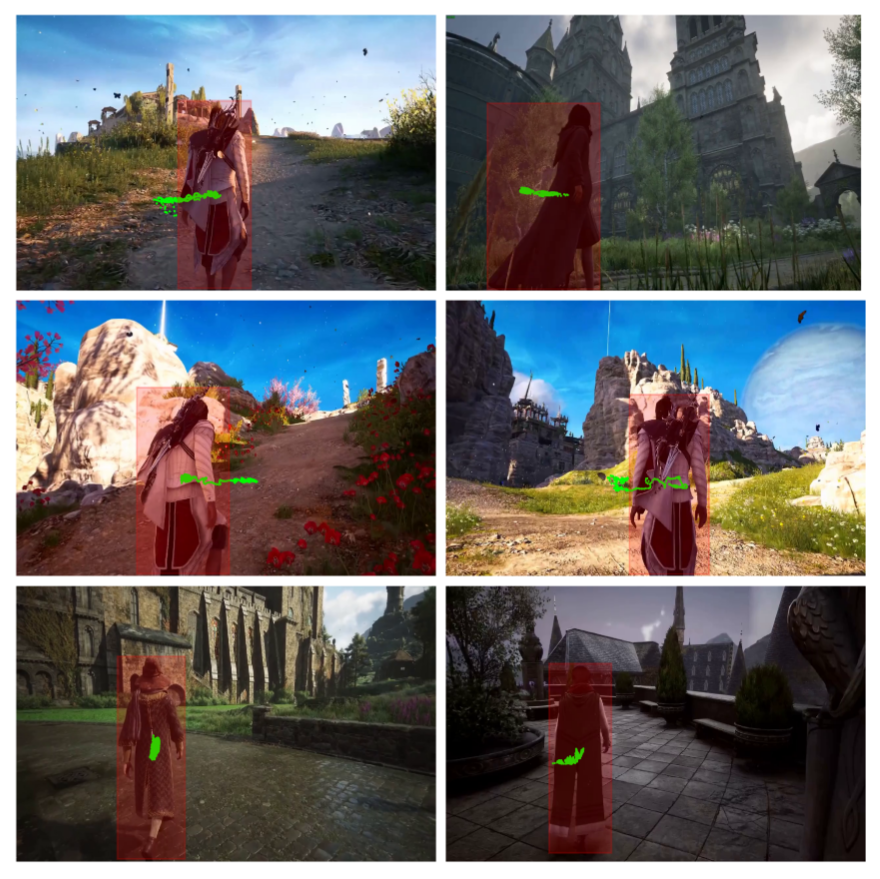}
    \caption{Examples from the \textit{Walking The Maps} dataset showing the bounding box of the subjects and their corresponding movement represented as trajectories of the center of the bounding box.}
    \label{fig:samples_example.pdf}
\end{figure}

\textit{Walking The Maps} comprises video clips extracted from publicly available YouTube walkthroughs of popular video games. These clips were selected because of the high quality audio and video in modern games, making them ideal sources for creating realistic V2A datasets. Footsteps sounds in these videos are often clearly audible and strongly characterized, with distinct acoustic profiles depending on the surface type (e.g., grass, concrete, wood) and movement dynamics (e.g., slow walking vs. running). Additionally, the high-definition visuals in these clips provide a robust temporal, semantic and spatial relationship between audio and video.

The dataset includes clips from four popular video games: \textit{Hogwarts Legacy}, \textit{Zelda Breath of the Wild}, \textit{Assassin's Creed: Odyssey}, \textit{Assassin's Creed IV Black Flag}.
It is composed of clips where the sound of footsteps is clearly audible and unaccompanied by other interfering sounds. This ensures a clean relationship between visual content and the corresponding audio. A total of 893 video clips were collected, with an average duration of 8.82 seconds. The shortest clip is 2.04 seconds, while the longest extends to 72.05 seconds.  
Each clip is labeled with metadata that includes the unique YouTube video ID, the start time, and the end time, resulting in filenames of the format $\mathrm{ID\_start\_end}$.mp4. The associated audio was processed using the AudioSep model \cite{liu22_winterspeech} with the textual query ``footstep sounds" to isolate the footsteps from background noise, resulting in clean, high-quality target audio. Some examples representing different characters with the related bounding boxes are available in Fig.~\ref{fig:samples_example.pdf}. 

Since our method is based on the use of foundation models employed in a zero-shot fashion, the use of a small dataset such as \textit{Walking The Maps} does not represent a constraint for our purposes, indeed it is a proof of the  generalization capability of the model to extract the information of interest.
\textit{Walking The Maps} provides a valuable resource for evaluating V2A models, especially for tasks requiring precise semantic, spatial, and temporal alignment between video and audio. 

\subsection{Evaluation Metrics}

To evaluate the quality of the proposed method, we employ different objective metrics that assess audio quality and spatial and temporal alignment. Below, we describe each metric in detail:

\begin{itemize}

\item Fréchet Audio Distance (FAD): FAD \cite{KilgourZRS19} metric measures the audio quality of generated samples by comparing their feature distribution to that of real audio samples. This metric focuses on evaluating the fidelity and realism of the generated audio. Lower FAD scores correspond to better-quality audio that is more perceptually similar to real-world sounds.
We use Laion-CLAP \cite{laionclap2023} encoder to generate embeddings from real and synthetic distributions on which calculate the FAD metric.

\item Fréchet Audio-Visual Distance (FAVD): FAVD \cite{10.1007/978-3-031-72986-7_17} metric is specifically designed to evaluate video-to-audio (V2A) models by measuring the alignment between audio and video in both temporal and semantic dimensions. It calculates the Fréchet Distance between the embeddings of video and audio modalities. For this purpose, we use I3D \cite{Carreira2017QuoVA} as the video encoder and VGGish \cite{Hershey2016CNNAF} as the audio encoder. By comparing embeddings extracted from ground truth videos and their corresponding generated audio, FAVD provides a quantitative measure of how well the generated audio aligns with the visual content. Lower FAVD scores indicate better alignment.

\item Spatial Audio-Visual Alignment (Spatial AV-Align): introduced in \cite{shimada2024savgbench}, this metric evaluates spatial synchronization between sound events in the generated audio and corresponding objects in the video. This metric ranges from 0 to 1, where higher values indicate better spatial alignment. The metric leverages object detection and sound event localization techniques.
\begin{figure*}
    \centering
    \includegraphics[width=\linewidth]{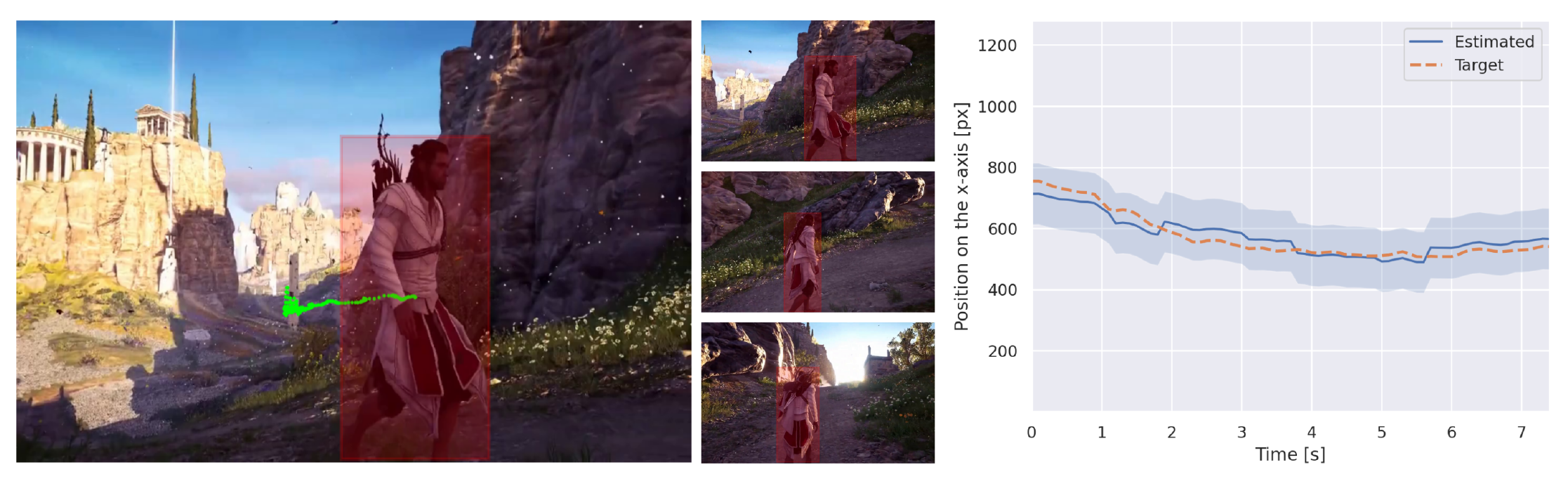}
    \caption{\textbf{Visual example illustrating the behavior of the Spatial AV-Align metric.}
On the left, the movement of the subject’s bounding box is shown as a series of green dots. In the center, a set of video frames displays the subject in various spatial positions. On the right, a plot shows with an orange dotted line the evolution over time of the center of the bounding box projected on the x-axis (target) and with a blue line the position (direction) on the x-axis over time of the sound as estimated by the SELD net. In light blue there is the In light blue is the confidence interval within which the detection is considered correct. Ideally, the trajectory of the sound direction and the bounding box center should align.}
    \label{fig:av-align}
\end{figure*}

1) \textit{Object Detection}: it uses YOLOX \cite{yolox2021} to detect objects in the video frames. The object detector identifies candidate object positions in each frame, processing videos at a frame rate of 4 \textit{fps}.

2) \textit{Sound Event Localization and Detection} (SELD): A stereo SELD model \cite{adavanne2018sound, wilkins2023two} is used to detect and localize sound events in the stereo audio. It determines the horizontal positions of sound sources in each 100 ms window of the audio.
For each sound event detected in the audio frame, the metric checks if it overlaps with a detected object in the nearest video frame. Overlap is considered a true positive ($TP$), while non-overlap is a false negative ($FN$). To account for temporal context, object detections from adjacent video frames are also considered valid. The Spatial AV-Align score is computed as:
\[
\text{Alignment Score} = \frac{TP}{TP + FN}
\]
where TP and FN represent the counts of true positives and false negatives, respectively.
These metrics indicate the spatial alignment of the generated audio-visual content.

\item E-L1: we measure how well the generated sound temporally matches the ground truth sound using the E-L1 introduced in \cite{Chung2024TFoleyAC}. This metric computes the L1 between the envelope extracted from the generated waveform and the original waveform:
\begin{equation}
        E\text{-}L1 = \frac{1}{k} \sum_{i=1}^{k} \|\mathbf{r}_i - \hat{\mathbf{r}}_i\|,
    \end{equation}
where $\mathbf{r}_i$ is the ground-truth envelope of the i-th audio frame, and $\hat{\mathbf{r}}_i$ is
the predicted one.
\end{itemize}

\subsection{Training Details} 

\begin{table*}[t!]
    \caption{Results for StereoSync and ground truth on the evaluation metrics.
    FAVD and FAD evaluate the overall audio quality, respect video and ground truth audio,
    while Spatial AV-Align metric indicates the spatial alignment between video and audio.}
    \centering
    \begin{tabular}{lcccc}
        \toprule
        \multirow{2}{*}{\textbf{Model}} & \multicolumn{1}{c}{\textbf{Spatial alignment}} & \multicolumn{2}{c}{\textbf{Semantic alignment}} & \multicolumn{1}{c}{\textbf{Temporal alignment}} \\
        \cmidrule(lr){2-2} \cmidrule(lr){3-4} \cmidrule(lr){5-5}
         & \textbf{AV-Align $\uparrow$} & \textbf{FAD-LC $\downarrow$}  & \textbf{FAVD $\downarrow$} & \textbf{E-L1 $\downarrow$} \\
        \midrule
        
        StereoSync                 & \textbf{0.78} & 0.230  & 3.8301 & 0.047 \\
        StereoSync (w/o spatial cond.) & 0.61 & 0.256 & 3.0682 & 0.062 \\
        \midrule
        Ground Truth               & 0.89 & - & - & - \\
        
        \bottomrule
    \end{tabular}
    \label{tab:evaluation_results}
\end{table*}

\label{sec-F:exp}

For training StereoSync, we initialize the model weights using the Stable Audio Open repository and its associated checkpoint. The ground truth audio used in our experiments is 44.1 kHz stereo recordings from the \textit{Walking The Maps} dataset. The model is trained on a single Nvidia RTX A6000 GPU (48 GB) with a batch size of 12 for only 3,000 steps. The training process employs the AdamW optimizer, with parameters configured as those in Stable Audio Open, and uses a fixed learning rate of \( 1 \times 10^{-4} \).  

MASA, RollingDepth and CLAP models are used zero-shot on our dataset, using official repositories and checkpoints for all of them. The generalization capabilities of these foundation models allowed us to avoid performing any fine-tuning on the dataset, being able to use these frameworks directly on \textit{Walking The Maps} videos and waveforms. 
In addition, generating depth maps and bounding boxes through deep learning models instead of classical algorithms for processing them, allowed us to exponentially reduce the time required for video pre-processing and spatial information extraction.\\
During inference, envelopes extracted directly from ground truth audio are interpolated to match the target sample rate, and fed into audio synthesis model's ControlNet as inputs. The model then generates the final output in 150 sampling steps, applying classifier-free guidance with a guidance scale set to 2.

\section{Results}


We compare the outputs of our method against the ground truth, and we also include a version of our model without spatial conditioning to underscore the benefits of this addition. Since, at the time of writing this paper, no other methods generate spatially aligned stereo audio from video, comparing our results directly to the ground truth remains the most effective way to evaluate our approach.
For spatial alignment, we rely on the Spatial AV-Align metric — which integrates SELD and object detection — and note that while the ground truth scores 0.89, StereoSync achieves 0.78. This indicates that our approach maintains spatial consistency in the stereo channels, whereas the variant without spatial cues only scores 0.61. Figure \ref{fig:av-align} gives a visual example of the prediction of sound direction and bounding box that are used by Spatial AV-Align for the purpose of metric calculation.\\
The FAD and FAVD metrics, which respectively evaluate the distance of the generated audio from the ground truth and the input video to be sonorized, highlight the ability of our model to generate high quality 44.1kHz stereo audio consistent with the required semantics. As StereoSync was not designed to improve semantic alignment compared to the non-spatially conditioned version, these results align with our expectations. In addition, both models are evaluated on temporal alignment using the E-L1 metric, confirming that the introduction of spatial conditioning does not compromise both the timing and quality of the generated audio.\\
Table \ref{tab:evaluation_results} summarizes these findings, demonstrating that while our primary goal is to enhance spatiality, the overall semantic and temporal alignment of the audio remains robust.

\label{sec:res}

\section{Conclusion}

In this work, we introduced StereoSync, a novel V2A framework leveraging foundation models for generating immersive and spatialized audio for silent videos. Our approach integrates MASA and Rolling Depth for video representation, alongside CLAP encoder and Stable Audio for realistic and high-quality sound generation. By leveraging the capabilities of these pretrained models, we enhance both the semantic and spatial coherence of the generated audio, ensuring alignment with the visual content while maintaining our model lightweight and fast.
To evaluate our method, we employed FAD and FAVD to assess the overall semantic quality of the generated waveforms. We then use the Spatial AV-Align metric, specifically designed to measure the spatial coherence between visual and audio modalities. Experimental results demonstrate that our stereo-based approach captures spatial relationships effectively, reinforcing the importance of spatial synchronization in generative audiovisual systems.

Although the Spatial AV-Align metric is useful for quantifying the spatial alignment between audio and video, we are aware that it suffers from a certain degree of error. In the case of \textit{Walking The Maps}, the sound action takes place in a central area of the frame. In this scenario, the metric may incorrectly assign better scores to a model that always places sounds to the center compared to one that attempts to place sounds according to the position of the subject in the video. In this paper, we have considered this undesirable behaviour and attempted to limit it through human supervision, however, as part of future work, we will focus on formulating a more effective metric. Furthermore, we will refine spatial representations, using different types of conditionings and losses in order to generate binaural audio and spatial audio with a higher number of channels.
\label{sec:con}

\section{Acknowledgements}
This work was supported by the European Union under the Italian National Recovery and Resilience Plan (NRRP) of NextGenerationEU, partnership on “National Centre for HPC, Big Data and Quantum Computing” (CN00000013 - Spoke 6: Multiscale Modelling \& Engineering Applications). 

This work was supported by “Progetti di Ricerca Medi” of Sapienza University of Rome for the project “SAID: Solving Audio Inverse problems with Diffusion models”, under grant number RM123188F75F8072.

\bibliography{ref}
\bibliographystyle{ieeetr}
\end{document}